# TOWARDS AN IO INTENSIVE GRID APPLICATION INSTRUMENTATION IN MEDIOGRID

**Dacian Tudor[1], Florin Pop[2], Valentin Cristea[2], Vladimir Cretu[1]**

[1] *Computer Science and Engineering Department, "Politehnica" University of Timisoara*
*{dacian, vcretu}@cs.utt.ro*
[2] *Faculty of Automatics and Computer Science, University "Politehnica" of Bucharest*
*{florinpop, valentin}@cs.pub.ro*

Abstract: Obtaining high performance in IO intensive applications requires systems that support reliable fast transfer, data replication, and caching. In this paper we present an architecture designed for supporting IO intensive applications in MedioGRID, a system for real-time processing of satellite images, operating in a Grid environment. The solution ensures that applications which are processing geographical data have uniform access to data and is based on continuous monitoring of the data transfers using Mon*ALISA* and its extensions. The MedioGRID architecture is also built on Globus, Condor and PBS and based on this middleware we aim to extract information about the running systems. The results obtained in testing MedioGRID system for large data transfers show that monitoring system provides a very good view of system evolution.

## 1. INTRODUCTION

An important feature that grid middleware provides to grid applications is uniform and location transparent access to data. In case of grid applications that intensively handle large amounts of data, identifying the bottlenecks and taking the appropriate measures to support this feature and ensure a smooth execution are very difficult. It requires the adoption of some services to monitor the operations at the data communication and execution levels and mechanisms to use these services transparently for the applications and other grid services. Grid middleware distributions like LCG or gLite have monitoring components that provide information about IO transfers. Mon*ALISA*, Ganglia, GridICE, NetFlow are monitoring components that can be included in grid middleware distributions or work independent in Grid systems.

In this paper we present the architecture of MedioGRID, a system for real-time processing of satellite images, operating in a Grid environment, and the monitoring solution for applications that doing the real-time processing of geographical data and have uniform access to data. The solution is based on continuous monitoring of the data transfers using Mon*ALISA* and its extensions.

The paper is structured as follows. Section 2 describes the MedioGRID system and its data handling concepts. Section 3 presents the monitoring IO intensive application highlighting efficiency related issues. In Section 4 the results obtained in experimental tests for data transfer with GridFTP between different sites are presented. The paper concludes with the presentation of further work.

## 2. DATA HANDLING IN MEDIOGRID

### 2.1 MedioGRID Project

The MedioGRID project (Ordean et al., 2005) aims to provide a grid based modern solution to early detection and prediction of water floods and fires by applying specialized transformation and processing on large satellite data sets. It is a collaborative project developed by several Romanian universities, located in different geographic areas and connected to the Romanian academic network RoEduNet. Each partner has separate physical resources (organized in clusters) they made available for the MedioGRID Virtual Organization. Different partner cluster can be linked together by VPN connections and use different distributions of Linux.

The Grid software infrastructure is provided by the Globus toolkit and a few grid add-on components like OGSA-DAI (see Figure 1). Applications in MedioGRID are developed using the client-server paradigm and web service interfaces.

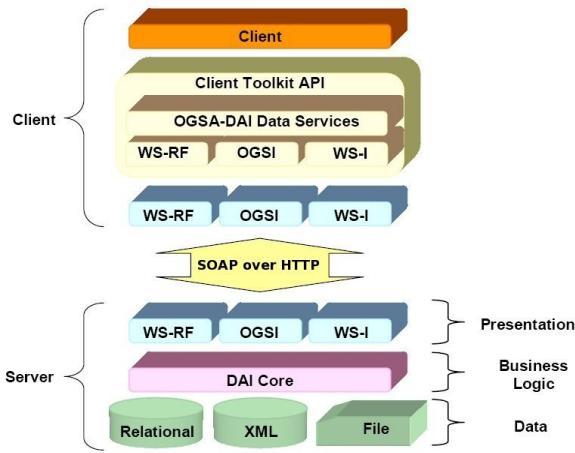

Figure 1. Client-Server model[1]

MedioGRID has a central image acquisition server, which receives the environmental data as a collection of MODIS (Modis, 2006) images. A MODIS image is a collection of files that include 36 channels covering a wide frequency spectrum, with a spatial resolution ranging from 250m to 1km. Each MODIS data corresponds to a geographic area. The server is expected to receive around 20 files per day, and per area which leads to an average of about 1000 files per day if we consider the number of areas for Romania.

Based on the stored environmental data, MedioGRID offers a set of public services that analyze the data, and apply different transformations. Considering the large data sets and the rich spectrum of data transformations and operations, MedioGRID has the characteristics of both data grids and computational grids, with a stronger orientation towards data sharing. Maintaining a good balance in terms of communication and computations requires tools and techniques to measure the activity and identify data access bottlenecks, take the necessary corrections, etc.

Applications in MedioGRID are IO intensive. Consequently, the analysis must be done with an IO intensive grid monitoring tool. This tool is one of the most important components of the MedioGRID architecture.

---

[1] Original from OGSA-DAI WS-RF User Guide, http://www.globus.org/toolkit/docs/development/3.9.5/techpreview/ogsadai/doc/background/JPGHighLevelArchitecture.jpg

*2.2 Data Replication*

Data replication in MedioGRID is static. Data is received by the central image acquisition server and is replicated to pre-defined machines, which are part of the grid. This type of replication has the advantage of a know-in-advance data traffic and ensures that, in any cluster, data is available for processing at any point in time.

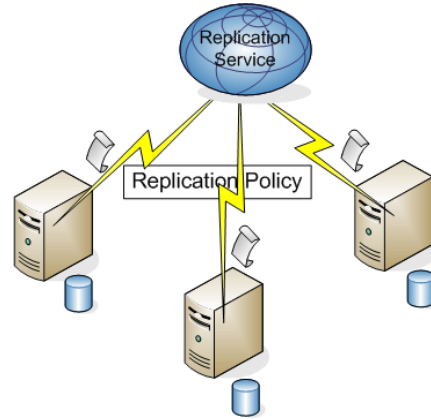

Figure 2. Data Replication

Traffic required by an operation from MedioGRID application is highly dependent on how many data sets must be retrieved and on the design of the data providing layer. By default the data retrieving layer simply locates the cluster-related data service and supplies the cluster located data. Using the OGSA-DAI interfaces we can have a uniform access to data. This policy could be altered to supply alternative data paths outside of the inner cluster in case there are many IO operations scheduled inside the cluster. Figure 2 describe a data replication system for MedioGRID. Replication servers is located in one cluster and have access to data using described policy. All data can be replicated on every node in a cluster.

*2.3 Data Access Patterns*

Grid data access patterns are determined by the static replication service running permanently and the requested data and services patterns. As replication patterns can be anticipated and act accordingly when designing the replication layer, the application dependent requests cannot be anticipated. Thus, in order to assess the current status, we would need a service or tool that is able to monitor IO intensive applications over the MedioGRID infrastructure.

The most simple access pattern is the local access. In this case the data is cached already at the node where data processing takes place. Thus, no further data transfers are required. If the data is not available and the entire data set is within the cluster, several transfers are required in order to complete the processing algorithm.

More complex interactions require operating on the intermediate data in a pipe fashion. If the computation node can handle all the intermediate operations, we have again no additional transfer. In a grid scenario, where different services are provided by different service deployed at different nodes, potentially outside of the organization boundaries, such interaction leads to a high and uncontrolled traffic pattern. Such a pattern cannot be a priori determined, but can be influenced by the service deployment. However, such a deployment configuration would be against the grid purpose, leading to a more likely limited system with fixed and predefined configuration and interactions.

Thus, we believe that highly dynamic and wide spread data processing interactions in IO intensive applications, where data is produced in intermediate steps, require advanced monitoring of the data providing layer across the entire grid.

## 3. MONITORING IO INTENSIVE APPLICATIONS IN MEDIOGRID

*3.1 System architecture*

The proposed system architecture is based on a monitoring system for IO intensive applications. This system, whose architecture is presented in Figure 3, is completely decentralized, based on replicated components and, consequently, fault-tolerant.

*3.1.1 Data server*

Data server or Grid storage is a general term for any server component to store data. Each storage node from the system can contain its own storage medium, microprocessor, indexing capability and management layer. Grid storage has some advantages by introducing a new level of fault-tolerance and redundancy, through the existence of multiple data paths between each pair of nodes. Thus it ensures that the storage grid can maintain optimum performance under conditions of fluctuating load and hence being scalable. If a new storage node is added, it can be automatically recognized by the rest of the grid. This reduces the need for expensive hardware upgrades and downtime.

*3.1.2 Replication system*

Computational Grids normally deal with large computational intensive problems on small data sets. In contrast, Data Grids mostly deal with large computational problems that in turn require evaluating and mining large amounts of data. Replication is regarded as one of the major optimization techniques for providing fast data access (W. Bell, 2002).

*3.1.3 Data scheduling*

Data-intensive applications require high performance storage subsystems. Parallel storage systems are widely used in today's clusters. A *greedy* I/O scheduling method is a good strategy for this problem. The main objective is to reduce the numbers of I/O operations and improve the performance of the whole storage system. (Xinrong, 2003). It is also possible to consider a heuristic method for data scheduling based on prediction of available bandwidth.

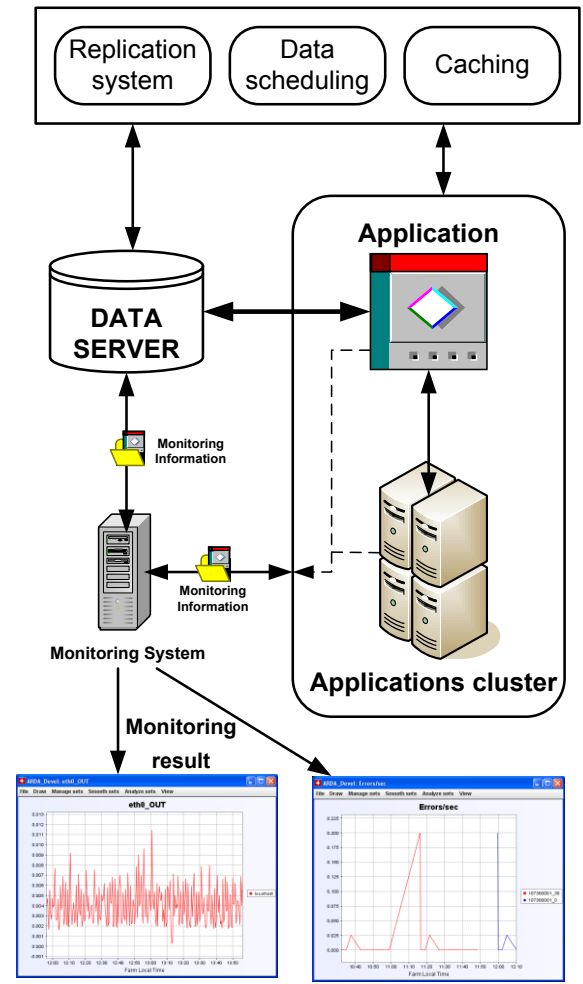

Figure 3. System architecture

*3.1.4 Monitoring system*

MedioGRID uses MonALISA for advanced monitoring. The MonALISA system is designed as an assembly of autonomous multi-threaded, self-describing agent-based subsystems which are registered as dynamic services and are able to collaborate and cooperate in performing a wide range of information gathering and processing tasks. MonALISA agents can analyze and process information in a distributed way, to provide optimized decisions in large scale distributed applications. An agent-based architecture provides the ability to invest the system with increasing degrees of intelligence, to reduce complexity and make global systems manageable in real time.

The scalability of the system derives from the use of multithreaded execution engine to host a variety of loosely coupled and self-describing dynamic services or agents and the ability of each service to register themselves and be discovered and used by any other services, or clients that require such information. The system is designed to easily integrate existing monitoring tools and procedures and to provide this information in a dynamic, customized, self describing way to any other services or clients.

MonALISA is currently running around the clock monitoring several Grids and distributed applications on around 160 sites (Mon*ALISA*, 2007). It has been developed over the last four years by Caltech and its partners with the support of the U.S. CMS software and computing program.

One of MonALISA extension is ApMon (Application Monitoring). The status of each of the computing nodes is monitored using ApMon which sends data to MonALISA services. A monitoring application built using the ApMon can send data about the status of a given application or about the status of the entire system on which the application runs. Java–based applications from MedioGRID have the purpose to periodically verify both the status of the monitored resource and of the tasks assigned to the computing resource (Mon*ALISA*, 2007).

*3.1.5 Monitoring results*

Obtaining monitoring results to provide accounting information is one the main scope of monitoring system. Many members of the same organization may run IO intensive jobs in a common user account on a machine or on a cluster. A job may fork a number of children jobs, which in turn may fork some children jobs *etc.*, so all the descendant jobs must be monitored. It is desirable to trace the entire path followed by a job from the user's computer to the machine on which it will finally be executed.

For example, Figure 4 presents the sum of ftp input traffic for one node in UPB MedioGRID cluster. This is one of the possibilities to view the monitoring results, time interval being short (upon 6 hours).

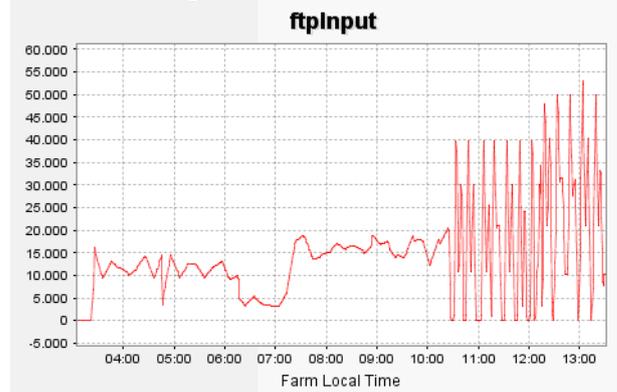

Figure 4. Total Ftp Input traffic for one UPB MedioGRID Node

Accounting information for input and output ftp traffic (provided by GridFTP service) for different VO is presented in Figure 5. For this method time interval could be larger, like one week, one month, three months, six months or one year.

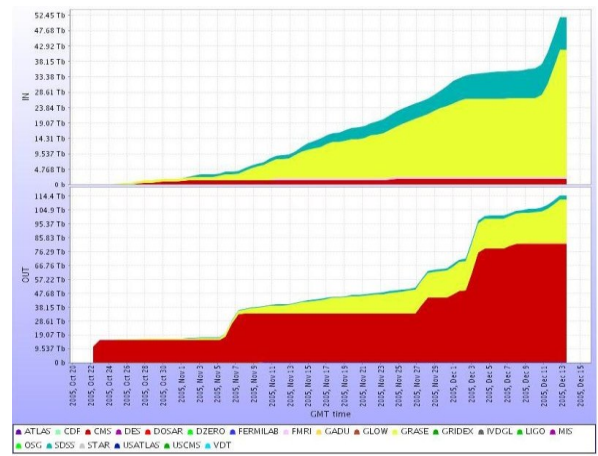

Figure 5. IO Accounting for different VO in Grid

Both of these figures are obtained with MonALISA which have modules that hold statistical information about the GridFTP traffic. The collected values are displayed in the MonALISA client or in the MonALISA repository like integrated values (accounting system).

*3.2 Monitoring application in MedioGRID*

MedioGRID project develops applications for satellite image processing. Extracting relevant environmental and meteorological parameters is another scope for this project. Applications for these are: Greenland (green zone detection) developed by UTCN and Change Detection developed by UPB. This Grid applications use registered services and tools like query, monitoring, discovery, security, registration, management, scheduling.

According with our presentation about GRID infrastructure and GRID monitoring solutions for MedioGRID (Muresan et al, 2006), we design an architecture for grid monitoring which is much easier if you know what to expect and which are the main work items. We plan to use a development environment or toolkit specially designed for grid applications, Globus Toolkit, MonALISA, Ganglia, ApMon and other middleware resources.

The MedioGRID solution combines MonaLISA and Ganglia. With Ganglia we have access to each node in cluster and we can request all information about state of node: load, CPU usage, memory consumption etc. In order to centralize this date on a single node (e.g. a server in the cluster) we use MonaLISA because we have support for collecting this type of data. We can collect information about jobs state or another kind of parameters about jobs, parameters created by users with ApMon and forwards them into MonaLISA database (MonALISA project, 2006). From MonaLISA database we can see the history about cluster in a repository.

*3.3 Interaction with Globus architecture*

The open source Globus Toolkit is one the fundamental enabling technology for the Grid systems, letting virtual organizations share computing power, databases, and other tools securely online across corporate, institutional, and geographic boundaries without sacrificing local autonomy. The toolkit includes software services and libraries for resource monitoring, discovery, and management, plus security and file management.

GridFTP is a high-performance, secure, reliable data transfer protocol optimized for high-bandwidth wide area networks. The GridFTP protocol is based on FTP, the highly-popular Internet file transfer protocol. GridFTP provides the protocol features such as: GSI security on control and data channels, multiple data channels for parallel transfers, partial file transfers, third-party (direct server-to-server) transfers, authenticated data channels, reusable data channels, command pipelining.

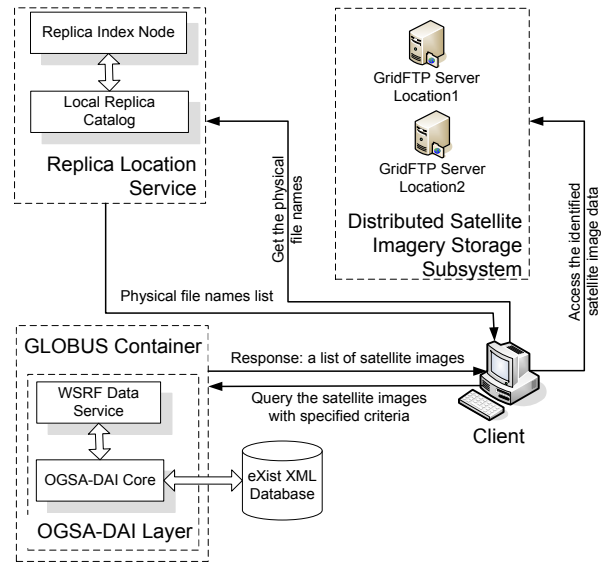

Figure 6. Accessing the Satellite Imagery Repository

Replica Location and Management supports multiple locations for the same file throughout the grid. Using the replica management functions, a file can be registered with the Replica Location Service (RLS) and its replicas can be created and deleted. Within RLS, a file is identified by its Logical File Name (LFN) and is registered within a logical collection. The record for a file points to its physical locations. This information is available from the RLS upon querying.

This two are non web services component from Globus. MedioGRID IO intensive applications are developed as web services using Globus architecture for data management, but application are base in the middleware level on GridFTP and Replica Location.

Figure 6 shows a very simple method for a client application to access the MedioGRID satellite imagery repository by using OGSA-DAI, XML databases, RLS and GridFTP. There is also possibility of creating a distributed storage network based on GridFTP, RLS and a central metadata repository.

4. EXPERIMENTAL RESULTS

The MedioGRID project is at the middle stage of development. At the current stage a dedicated network architecture connecting all the participants has been implemented and some applications are started. All the data transfers in MedioGRID are performed using the Globus GridFTP protocol. It provides means of tweaking several performance parameters which directly influence the overall data transfer performance. One of these parameters is the GridFTP parallelism degree (number of parallel data connections used at a time) and the file size.

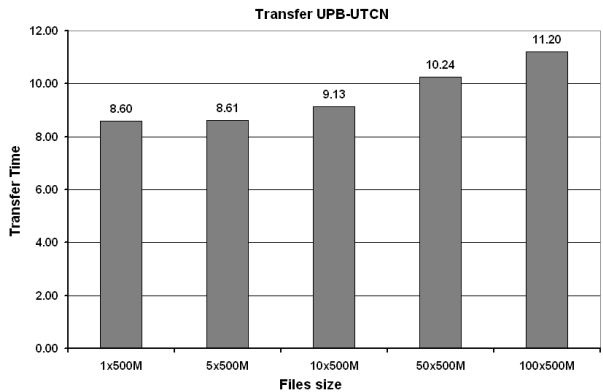

Figure 7. Experiment 1: FTP transfer inter-cluster

The Figure 7 presents the transfer time in seconds between the storage server and a high speed client at the MedioGRID system. Four 500Mb data sets with different file sizes (1x500Mb, 5x100Mb, 10x10Mb, 50x10Mb, 100x5Mb) have been used to perform one of the test for an application developed in UPB.

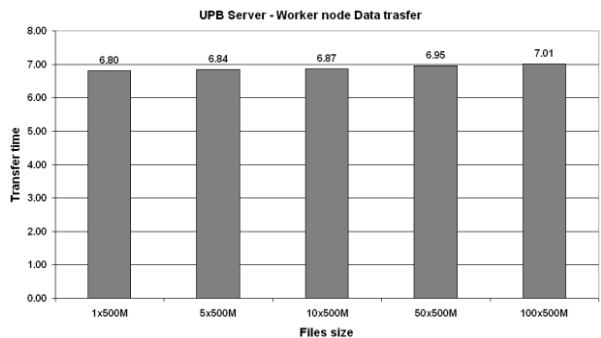

Figure 8. Experiment 2: FTP transfer intra-cluster

This application (dedicated for change detection in satellite image) transfer data from UTCN data storage server and then transfers this data between nodes cluster in UPB. This graphic demonstrates that the optimal performance is achieved when using a number less than 15 GridFTP parallel data streams.

The intra-cluster data transfer characteristics in MedioGRID have been measured using transfers between data processing clients located at the UPB site. The same data like in Experiment no. 1 have been used. Graphic from Figure 8 demonstrate that the data transfer performance is directly influenced by the number of parallel data streams. The graphic demonstrates that, the transfer time is the same for all data from experimental set.

## 5. CONCLUSIONS AND FUTURE WORK

In this paper we propose a high level architectural specification of a monitoring system for IO intensive applications in MedioGRID project. One of the important results in monitoring Grid environment and instrumentation application is obtaining accounting information. The accounting information can be used in intelligent scheduling of data transfer, involves collecting statistics on the types and status of transfers ordered by users, transfer rate and consumption of bandwidth.

We include in this system a monitoring tools and extension for collecting data from application running on a cluster and from cluster. We propose MonALISA system and its extension because it is a decentralized monitoring system and interact directly with Globus. We also present the possibility to view the monitoring data.

We test the system using a simple application for some large transfers and obtain an optimum for number of parallel data streams. Next phase of our research will approach a real-time IO intensive application and provide a complex accounting system.